\documentclass{article}
\usepackage{quel-art}
\usepackage{amsmath}
\newcommand{\ket}[1]{|#1\rangle}
\newcommand{\bra}[1]{\langle#1|}

\volnumber{} \issuenumber{} \edyear{}                             
\frompage{} \topage{}                                              
\recrevdate{}                                                            
\def\rightmark{On geometric phases for quantum trajectories}
\def\leftmark{E. Sj\"oqvist} 
\title{On geometric phases for quantum trajectories}
\authors{ 
{Erik Sj\"oqvist$^{1,a}$  
\index{Sj\"oqvist, E.} 
\index{Two, A.} 
}\\[2.812mm] 
{\normalsize 
\hspace*{-8pt}$^1$ Department of Quantum Chemistry, \\  
Uppsala University, Box 518, Se-751 20 Uppsala, Sweden}} 
  
\abstract{A sequence of completely positive maps can be decomposed 
into quantum trajectories. The geometric phase or holonomy of such a
trajectory is delineated. For nonpure initial states, it is shown that
well-defined holonomies can be assigned by using Uhlmann's concept of
parallel transport along the individual trajectories. We put forward
an experimental realization of the geometric phase for a quantum
trajectory in interferometry. We argue that the average over the phase
factors for all quantum trajectories that build up a given open system
evolution, fails to reflect the geometry of the open system evolution
itself.}
\keyword{Geometric phase, open systems, quantum trajectories} 
 
\PACS{03.65.Vf, 03.65.Yz} 
  
\makeindex 
\begin{document} 
  
\maketitle 
 
\section{Introduction}
\label{intro} 
Imagine a quantum particle prepared in a pure state. If the state
evolves in a unitary fashion so that it remains pure, this particle 
is said to constitute a closed system. Berry \cite{berry84} and 
others \cite{aharonov87,samuel88,mukunda93} have shown that a closed system 
picks up a phase factor of geometric origin; a geometric phase 
factor, or quantum holonomy.

But what happens if the system fails to be closed? Is there a natural
generalization of the geometric phase to open systems? Considerations
of phases of geometric origin for open quantum systems go back to work
by Uhlmann \cite{uhlmann86} on holonomy accompanying density
operators. Garrison and Wright \cite{garrison88} have assigned
complex-valued geometric phases to open systems on a phenomenological
basis by the use of non-Hermitian Hamiltonians. Ellinas {\it et al.} 
\cite{ellinas89} have introduced geometric phases associated with
adiabatic evolution of eigenmatrices of the Liouvillian superoperator
describing an open system. Gamliel and Freed \cite{gamliel89}
demonstrated that the adiabatic Berry phases may appear also in
the presence of open system effects, as phase factors in the
off-diagonal elements of the density operator expressed in the
instantaneous energy eigenbasis (see also
\cite{fonseca02,kamleitner04}).

More recently, open system geometric phases have been
revisited from various perspectives, such as interferometry
\cite{ericsson03,peixoto03,tong04}, quantum trajectories 
\cite{carollo03,fuentes05}, phase distributions \cite{marzlin04}, 
the adiabatic theorem \cite{sarandy06}, and stochastic Schr\"odinger 
evolutions \cite{bassi06}. A reason for this renewed interest is 
the need to better understand the robustness properties of geometric 
phases in the presence of open system effects, triggered by the conjectured 
\cite{zanardi99} importance of quantum holonomy for fault tolerant 
quantum computation. To address the robustness issue, it is commonly
assumed that the system evolves under a Markovian master equation. 
The quantum trajectory approach is specifically designed for the 
analysis of the robustness of geometric phases to Markovian open 
system effects (see, e.g., \cite{pachos04,cen04}).

In this paper, we focus on the quantum trajectory approach to the open
system geometric phase. The purpose is fourfold. First, we wish to
reformulate the quantum trajectory approach to the geometric phase on
the basis of discrete sets of completely positive maps (CPMs), without
invoking continuous master equations. This reformulation is a
generalization of previous treatments \cite{carollo03,fuentes05} 
(see also \cite{carollo05}) in the sense that it yields the master
equation in the continuous limit. Secondly, we wish to extend
\cite{carollo03} to quantum trajectories that start in a nonpure
state. To this end, it seems most natural to use Uhlmann's concept 
of parallel transport \cite{uhlmann86} along the trajectories.  Thirdly,
we put forward an experimental procedure to implement the geometric phase
for an individual trajectory. Finally, we briefly address whether the
geometric phases for individual quantum trajectories can be used to
define a meaningful geometric phase for an open system. The paper ends
with the conclusions.

\section{Open quantum systems and quantum trajectories}
Consider a quantum system $s$ in contact with some environment $e$,
also of quantum nature. Suppose $s+e$ is in the state $\varrho$. Any
observable property pertaining only to $s$ is given by the reduced
density operator $\rho$ obtained by tracing $\varrho$ over $e$. 
Let $U(t,t_0)$ be the time evolution operator for $s+e$. Further, 
let $\ket{e_0},\ldots,\ket{e_{\mu}}$ ($\mu$ could be finite 
or infinite) be a complete basis for $e$ and suppose that $\varrho = 
\ket{e_0} \bra{e_0} \otimes \rho$ is prepared at $t_0$. Then, for 
an arbitrary $t$, the state of $s$ reads
\begin{eqnarray} 
\rho (t) = \mathcal{E}_{t,t_0} (\rho) = 
\sum_{p=0}^{\mu} E_p (t,t_0) \rho E_p^{\dagger} (t,t_0) ,  
\label{eq:cpm}
\end{eqnarray} 
where $E_p (t,t_0) = \bra{e_p} U(t,t_0) \ket{e_0}$ are the Kraus
operators \cite{kraus83} that constitute a Kraus representation of
$\mathcal{E}_{t,t_0}$. The map $\mathcal{E}_{t,t_0}$ is trace
preserving and completely positive, i.e., it takes normalized density
operators into normalized density operators and all trivial extensions
$\hat{1} \otimes \mathcal{E}_{t,t_0}$ likewise.

To introduce the concept of quantum trajectories it is convenient to
consider $N$ identical copies of the environment. Prepare them in the 
product state $\ket{e_0} = \ket{e_0^0} \otimes \ldots
\otimes \ket{e_0^{N-1}}$, with the superscript labeling the copies. 
Let $s$ interact only with the first copy between $t_0$ and $t_1$, 
only with the second copy between $t_1$ and $t_2$, and so on. These 
interactions are described by the unitary operators $U^0(t,t_0), 
\ldots,$ $U^{N-1} (t,t_{N-1})$, respectively. It results 
in the composite map
\begin{eqnarray}
\rho \rightarrow \mathcal{E} (\rho) = \mathcal{E}_{t_N,t_{N-1}} \circ 
\ldots \circ \mathcal{E}_{t_1,t_0} (\rho) .
\label{eq:compmap} 
\end{eqnarray}
Here, $E_p (t_{k+1},t_k) = \bra{e_p^k} U^k(t_{k+1},t_k) \ket{e_0^k}$,
$p=0,\ldots,\mu$, can be taken as Kraus operators for
$\mathcal{E}_{t_{k+1},t_k}$, $k=0,\ldots,N-1$. In terms of the above
$s+e$ picture, the relevant part of the input state at $t_k$ is
$\ket{e_0^k}\bra{e_0^k}\otimes \rho_k$, where $\rho_0 = \rho$ and 
$\rho_{k\in [1,N-1]} = \mathcal{E}_{t_k,t_{k-1}} \circ \ldots 
\circ \mathcal{E}_{t_1,t_0} (\rho)$. A quantum trajectory $\alpha$ 
is a sequence of (unnormalized) states $\rho \rightarrow 
\rho_1^{\alpha} \rightarrow \ldots \rightarrow \rho_N^{\alpha}$, where   
\begin{eqnarray} 
\rho_k^{\alpha} = E_{\alpha(k)} (t_k,t_{k-1}) \ldots E_{\alpha(1)} 
(t_1,t_0) \rho E_{\alpha(1)}^{\dagger} (t_1,t_0) \ldots 
E_{\alpha(k)}^{\dagger} (t_k,t_{k-1}),
\label{eq:states} 
\end{eqnarray}
for $k=1,\ldots,N$. Here, $\alpha (l) \in [0,\mu]$ is the l$th$
element of a sequence of indexes. The map $\mathcal{E}$ is recovered
by summing over all trajectories.

\section{Geometric phase of quantum trajectories}
\label{sec:gp} 
There is in general an infinite number of possible Kraus
representations for a given CPM. For the composite map $\mathcal{E}$
in Eq. (\ref{eq:compmap}), different choices of Kraus representations
in each step lead to different sets of trajectories. It follows that
no physical meaning can be associated to an individual trajectory
without imposing some additional physical constraint.

Nevertheless, it is still possible to formally associate a geometric
phase or holonomy to an individual quantum trajectory. Here, we wish
to do so. In particular, we wish to stress the importance of whether
the initial state of the trajectory is pure or nonpure. In the pure
case, which has been treated in \cite{carollo03,carollo05}, 
the standard Pancharatnam connection can be used, while in the nonpure
case we need to address the issue of parallel transport of density
operators.
 
Let us first consider the case where the initial state is pure, i.e., 
$\rho = \ket{\psi} \bra{\psi}$. Here, the trajectories may be 
lifted to sequences of (unnormalized) vectors: $\ket{\psi} \rightarrow 
\ket{\psi_1^{\alpha}} \rightarrow \ldots \rightarrow \ket{\psi_N^{\alpha}}$, 
where $\ket{\psi_k^{\alpha}} = E_{\alpha(k)} (t_k,t_{k-1}) \ldots 
E_{\alpha(1)} (t_1,t_0) \ket{\psi}$. The Pancharatnam connection 
\cite{pancharatnam56} yields the geometric phase factor of the 
trajectory $\alpha$ as 
\begin{eqnarray}
\gamma^{\alpha} = 
\Phi \big[ \bra{\psi} \psi_N^{\alpha} \rangle \bra{\psi_N^{\alpha}} 
\psi_{N-1}^{\alpha} \rangle \ldots \bra{\psi_1^{\alpha}} \psi \rangle 
\big] , 
\label{eq:pure}
\end{eqnarray} 
where $\Phi \big[ z \big] = z/|z|$ for any nonzero complex number $z$. 

When the initial state is nonpure, though, one needs to consider
geometric phases for mixed states, since the trajectories then contain
nonpure states $\rho,\rho_1^{\alpha},\ldots,\rho_N^{\alpha}$ (see
Eq. (\ref{eq:states})). To deal with this, it seems most natural to
use the Uhlmann approach \cite{uhlmann86}, since this approach is
particularly well adopted to discrete sequences of density
operators. Assume that $\rho,\rho_1^{\alpha},\ldots,\rho_N^{\alpha}$
are full rank and introduce the amplitudes $W=\sqrt{\rho}V$ and
$W_k^{\alpha} = \sqrt{\rho_k^{\alpha}} V_k^{\alpha}$, $k=1,\ldots,N$,
where $V,V_1^{\alpha},\ldots,V_N^{\alpha}$ are the unitary operators
or `phases' of the amplitudes. For a given but arbitrary $V$,
$V_1^{\alpha}$ is fixed by the parallelity condition
$W_1^{{\alpha}\dagger} W > 0$ and similarly $V_{k+1}^{\alpha}$ is
fixed iteratively by $W_{k+1}^{{\alpha}\dagger} W_k^{\alpha} > 0$, for
$k=1,\ldots,N-1$. In this way, the final unitary operator
$V_N^{\alpha}$ is uniquely determined by the trajectory $\alpha$ up to
the arbitrary phase $V$ of the initial amplitude $W$. Explicitly,
\begin{eqnarray} 
V_N^{\alpha} & = & \Big( \sqrt{\rho_N^{\alpha}} \rho_{N-1}^{\alpha} 
\sqrt{\rho_N^{\alpha}} \Big)^{-1/2} \sqrt{\rho_N^{\alpha}} 
\sqrt{\rho_{N-1}^{\alpha}} \ldots 
\nonumber \\ 
 & & \times \Big( \sqrt{\rho_2^{\alpha}} 
\rho_1^{\alpha} 
\sqrt{\rho_2^{\alpha}} \Big)^{-1/2} \sqrt{\rho_2^{\alpha}} 
\sqrt{\rho_1^{\alpha}} \Big( \sqrt{\rho_1^{\alpha}} 
\rho \sqrt{\rho_1^{\alpha}} \Big)^{-1/2} \sqrt{\rho_1^{\alpha}} 
\sqrt{\rho} \, V .  
\end{eqnarray} 
To remove the arbitrary phase $V$, define 
\begin{eqnarray}
U^{\alpha} = V_N^{\alpha} V^{\dagger} ,  
\end{eqnarray}  
which is the holonomy of the trajectory $\alpha$. Note that 
$U^{\alpha}$ reduces to $\gamma^{\alpha}$ in the limit of 
pure initial states. 

\section{Accessing the geometric phase of a quantum trajectory}  
\label{sec:access}
As already mentioned, no physical meaning can be associated with a
single trajectory and its concomitant geometric phase or holonomy if
no further physical constraint is imposed. This is so because there
are infinitely many Kraus representations of a given CPM, and
therefore infinitely many equivalent ways to decompose the open system
evolution into quantum trajectories. Nevertheless, the geometric phase
for a single quantum trajectory can be accessed in principle by
performing measurements on the environment.

To see this, let us first make some simplifying assumptions. Suppose
that $\rho = \ket{\psi} \bra{\psi}$ (pure initial state) and that the
observables $O_e^k$, $k=0,\ldots,N-1$, with eigenstates
$\ket{e_0^k},\ldots,\ket{e_{\mu}^k}$, are measured projectively. We
further assume that these measurements are performed precisely at
$t_1,\ldots,t_N$, i.e., that the first copy of the environment is
measured at $t_1$, the second at $t_2$, and so on. In the first step,
$s+e$ evolves as (ignoring all copies except the first one)
\begin{eqnarray} 
\ket{e_0^0} \bra{e_0^0} \otimes \ket{\psi} \bra{\psi} & \rightarrow &  
U^0(t_1,t_0) \ket{e_0^0} \bra{e_0^0} \otimes \ket{\psi} \bra{\psi} 
U^{0\dagger}(t_1,t_0)
\nonumber \\ 
 & \rightarrow & \ket{e_{\alpha (1)}^0} \bra{e_{\alpha (1)}^0} \otimes 
E_{\alpha (1)} (t_1,t_0) \ket{\psi} \bra{\psi} E_{\alpha (1)}^{\dagger} 
(t_1,t_0) 
\label{eq:1st}   
\end{eqnarray}
for the registered outcome $\alpha (1) \in [0,\mu]$. In the second 
step, the state of $s+e$ similarly evolves as (now ignoring 
all copies except the second one)
\begin{eqnarray} 
 & & \ket{e_0^1} \bra{e_0^1} \otimes E_{\alpha (1)} (t_1,t_0) \ket{\psi} 
\bra{\psi} E_{\alpha (1)}^{\dagger} (t_1,t_0) 
\nonumber \\ 
 & & \rightarrow 
U^1(t_2,t_1) \ket{e_0^1} \bra{e_0^1} \otimes E_{\alpha (1)} 
(t_1,t_0) \ket{\psi} \bra{\psi} E_{\alpha (1)}^{\dagger} (t_1,t_0) 
U^{1\dagger} (t_2,t_1) 
\nonumber \\ 
 & & \rightarrow \ket{e_{\alpha (2)}^1} \bra{e_{\alpha (2)}^1} 
\otimes E_{\alpha (2)} (t_2,t_1) E_{\alpha (1)} (t_1,t_0) \ket{\psi} 
\bra{\psi} E_{\alpha (1)}^{\dagger} (t_1,t_0) 
E_{\alpha (2)}^{\dagger} (t_2,t_1) \hskip 0.5 cm 
\label{eq:2nd}   
\end{eqnarray}
for the registered outcome $\alpha (2) \in [0,\mu]$. Continuing in this 
way provides a physical realization of the trajectory $\alpha$. 

We now show that this measurement procedure can be used to
experimentally implement $\gamma^{\alpha}$. Suppose $\ket{\psi}
\bra{\psi}$ is the input state of the internal degree of freedom
(e.g. spin) of an ensemble of particles in an ordinary two-beam 
Mach-Zehnder interferometer. The evolution in Eq. (\ref{eq:1st}) 
is implemented in one of the interferometer arms by post-selecting 
those particles for which $\alpha(1)$ is registered. A variable 
U(1) phase shift is applied to the other arm. The maximum of the 
resulting interference fringes
is obtained for $\Phi \big[ \bra{\psi} E_{\alpha (1)}^{\dagger}
(t_1,t_0) \ket{\psi} \big] =
\Phi \big[ \bra{\psi_1^{\alpha}} \psi \rangle \big]$. 
In the next step, the internal state $E_{\alpha (1)} (t_1,t_0) \ket{\psi} 
\bra{\psi} E_{\alpha (1)}^{\dagger} (t_1,t_0)$ is used as input 
to a second interferometer in which Eq. (\ref{eq:2nd}) is implemented
in one arm by post-selecting the outcome $\alpha (2)$. The other arm
is again exposed to a variable U(1) shift, yielding an interference
maximum at $\Phi \big[
\bra{\psi} E_{\alpha (1)}^{\dagger} (t_1,t_0) E_{\alpha (2)}^{\dagger}
(t_2,t_1) E_{\alpha (1)} (t_1,t_0)\ket{\psi} \big] = \Phi \big[
\bra{\psi_2^{\alpha}} \psi_1^{\alpha} \rangle \big]$. Continuing 
in this way up to $\psi_N^{\alpha}$ and back to $\psi$ yields 
the phase factors $\Phi \big[ \bra{\psi} \psi_N^{\alpha} \rangle \big], 
\Phi \big[ \bra{\psi_N^{\alpha}} \psi_{N-1}^{\alpha} \rangle \big],$ 
$\ldots ,\Phi \big[ \bra{\psi_1^{\alpha}} \psi \rangle \big]$. 
By taking the product of these phase factors, we obtain  
\begin{eqnarray} 
\Phi \big[ \bra{\psi} \psi_N^{\alpha} \rangle \big] \Phi \big[ 
\bra{\psi_N^{\alpha}} \psi_{N-1}^{\alpha} \rangle \big] \ldots 
\Phi \big[ \bra{\psi_1^{\alpha}} \psi \rangle \big] = 
\Phi \big[ \bra{\psi} \psi_N^{\alpha} \rangle  
\bra{\psi_N^{\alpha}} \psi_{N-1}^{\alpha} \rangle \ldots 
\bra{\psi_1^{\alpha}} \psi \rangle \big] , 
\end{eqnarray}
which is precisely the geometric phase factor $\gamma^{\alpha}$ 
in Eq. (\ref{eq:pure}). It should be noted, though, that in order 
for this procedure to work it is in each step necessary to erase 
any path information related to the state change of the environment 
caused by the measurement. 

\section{Open system geometric phase} 
\label{sec:open}
So far, we have discussed how a geometric phase factor
$\gamma^{\alpha}$ or holonomy $U^{\alpha}$ can be assigned to a single
quantum trajectory $\alpha$ and how $\gamma^{\alpha}$ can be verified
in interferometry. However, since an individual quantum trajectory can
differ very much from the original evolution $\mathcal{E}$, it is not
obvious in what way $\gamma^{\alpha}$ gives information about the
geometry of the path that is generated by $\mathcal{E}$ itself. Such 
a connection can be established if it is possible to use the geometric 
phases for the trajectories to assign a well-defined geometric phase 
for $\mathcal{E}$. Here, we briefly address this issue.

The map $\mathcal{E}$ is recovered by summing over all trajectories
$\alpha$. In the case of pure initial state $\ket{\psi}$, it is
therefore natural to guess that there is a geometric phase factor
$\Gamma_{\mathcal{E}}$ of $\mathcal{E}$ that is the average over the 
phase factors of the individual quantum trajectories. Explicitly,  
\begin{eqnarray} 
\Gamma_{\mathcal{E}} = \sum_{\alpha} p_{\alpha} \gamma^{\alpha} , 
\end{eqnarray}
where $p_{\alpha} = \langle \psi_N^{\alpha} \ket{\psi_N^{\alpha}}$ is
the probability of $\alpha$. Unfortunately, though, this quantity
depends on the specific decomposition of $\mathcal{E}$ into
trajectories. Thus, unless a specific set of trajectories is singled
out (e.g., by repeatedly measuring the environment, as described in
the preceding section), no physical meaning can be assigned to
$\Gamma_{\mathcal{E}}$. An analogous conclusion in the context of
stochastic Sch\"odinger equations has been drawn in \cite{bassi06}.

\section{Conclusions}
\label{concl} 
The quantum trajectory approach has recently been proposed to tackle
the issue of geometric phase \cite{carollo03,carollo05} and quantum
holonomy \cite{fuentes05} for an open quantum system. The main merit
of the approach appears to be its usefulness in the analysis of the
robustness of geometric phase based quantum gates to open system
effects \cite{pachos04,cen04}. Here, we have reformulated the quantum
trajectory approach to the geometric phase, on the basis of sequences
of completely positive maps.

It has been stressed \cite{carollo03,carollo05} that the quantum
trajectory approach is useful in that it avoids the apparent need to
introduce parallel transport of nonpure states. This holds only under
the assumption that the trajectory starts in a pure state. We have
pointed out that even in the case where the initial state is nonpure
it is possible to assign well-defined holonomies by using Uhlmann's 
concept of parallel transport \cite{uhlmann86} along the quantum 
trajectories.

The geometric phase of an individual trajectory can be made physical
by repeatedly measuring the environment of the system. This opens up
the possibility to experimentally implement such phases by measuring
post-selectively the environment in one of the arms of a standard
Mach-Zehnder interferometry. In this way, we have put forward an
iterative procedure to measure the geometric phase of a single quantum
trajectory.

Finally, we have addressed the question whether the geometric phases
for the trajectories comprising an open system evolution can be used
to define the geometric phase of the open system evolution itself. A
natural approach is to sum over the individual phase factors weighted
by the corresponding probabilities. However, this works only if a
specific set of trajectories is singled out by some physical means. It
is interesting to note the analogous situation for the geometric phase
proposed in \cite{ericsson03}. This phase depends on the additional
detailed knowledge of the system-environment interaction. Thus,
neither the approach in \cite{ericsson03} nor the averaging over
quantum trajectories put forward here, result in a genuine concept of
geometric phase for an open system. It seems more appropriate to use
the path of density operator itself as the basis for such a concept
\cite{uhlmann86,tong04}.
   
\section*{Acknowledgment(s)}  
This work was financed in part by the Swedish Research Council. 
  
\section*{Notes} 
\begin{notes} 
\item[a] 
E-mail: eriks@kvac.uu.se 
\end{notes}

\vfill\eject 
\end{document}